\documentclass[reprint,amsmath,amssymb,aip,eqsecnum,jcp]{revtex4-1}
\usepackage[dvips]{graphicx}
\usepackage{color}
\usepackage{amsmath}
\usepackage{amssymb}
\usepackage{pstricks,pst-grad}
\usepackage{graphicx}
\usepackage{amsbsy}

\usepackage[linktoc=all,unicode=true,pdfusetitle,  bookmarks=true,bookmarksnumbered=false,bookmarksopen=false, breaklinks=true,pdfborder={0 0 0},backref=false,colorlinks=true]{hyperref}
\hypersetup{linkcolor=blue,citecolor=red}

\usepackage[hyphenbreaks]{breakurl}

\usepackage[normalem]{ulem}

\usepackage{tikz}
\usepackage{dcolumn}
\usepackage{bm}

\newcommand\beq{\begin{equation}}
\newcommand\eeq{\end{equation}}
\newcommand\beqa{\begin{eqnarray}}
\newcommand\eeqa{\end{eqnarray}}

\newcommand{\nn}{\nonumber\\}

\def\bal#1\eal{\begin{align}#1\end{align}}

\newcommand{\ex}{\text{ex}}

\begin{document}



\title{Virial coefficients and demixing in the Asakura--Oosawa model}


\author{Mariano L\'{o}pez de Haro}
\email{malopez@unam.mx}
\homepage{http://xml.ier.unam.mx/xml/tc/ft/mlh/}
\affiliation{Instituto de Energ\'{\i}as Renovables, Universidad Nacional Aut\'onoma de M\'exico (U.N.A.M.),
Temixco, Morelos 62580, M{e}xico}
\author{Carlos F. Tejero}
\email{cftejero@fis.ucm.es}
\affiliation{Facultad de Ciencias F\'{\i}sicas, Universidad Complutense de
Madrid,  E-28040 Madrid, Spain}
\author{Andr\'es Santos}
\email{andres@unex.es}
\homepage{http://www.unex.es/eweb/fisteor/andres/}
\author{Santos B. Yuste}
\email{santos@unex.es}
\homepage{http://www.unex.es/eweb/fisteor/santos/}
\affiliation{Departamento de F\'{\i}sica and Instituto de Computaci\'on Cient\'ifica Avanzada (ICCAEx), Universidad de
Extremadura,  E-06071 Badajoz, Spain}
\author{Giacomo Fiumara}
\email{giacomo.fiumara@unime.it}
\affiliation{
Department of Mathematics and Computer Science, University of Messina,
Viale F. Stagno D'Alcontres 31, I-98166 Messina, Italy}
\author{Franz Saija}
\email{franz.saija@cnr.it}
\affiliation{CNR-IPCF, Viale F. Stagno d'Alcontres, 37-98158 Messina, Italy}
\date{\today}

\begin{abstract}
The problem of demixing in the Asakura--Oosawa colloid-polymer model is considered. The critical constants are computed using truncated virial expansions up to fifth order.  While the exact analytical results for the second and third virial coefficients are known for any size ratio, analytical results for the fourth virial coefficient are provided here, and fifth virial coefficients are obtained numerically for particular size ratios using standard Monte Carlo techniques. We have computed the critical constants by successively considering the truncated virial series up to the second, third, fourth, and fifth virial coefficients. The results for the critical colloid and (reservoir) polymer packing fractions are compared with those that follow from available Monte Carlo simulations in the grand canonical ensemble. Limitations and perspectives of this approach are pointed out.
\end{abstract}

\maketitle

\section{Introduction}
\label{Intro}

Characterizing the phase behavior of a soft matter complex fluid, such as a colloidal suspension, is in general a very difficult task. This is  due to the widely different time and length scales involved. Therefore, the analysis of simple but tractable models capturing the essential features of real systems has proven to be a very valuable tool in this regard. Since the structure of a dense fluid is known to be largely
determined by the repulsive intermolecular forces, hard-core potentials have been extensively employed to model simple fluids and fluid mixtures. In the case of mixtures of colloids and polymers and other colloidal systems, binary hard-sphere {(HS)} mixtures are widely recognized as standard models for such systems.\cite{M08} A particularly interesting problem from the theoretical point of view is the possible existence of fluid-fluid separation in HS mixtures. The origin of a demixing transition in fluid mixtures is usually ascribed to the asymmetry of the interactions between the different components of the mixture. When one deals with a binary additive hard-sphere {(AHS)} mixture, the only possible asymmetry is that due to the different sizes of the spheres of both components. This means that the fluid-fluid separation that may occur in such a mixture, based only on the size asymmetry of the spheres, is entropically driven, an issue that has attracted a lot of attention in the literature.\cite{F94} A plausible mechanism for demixing in these mixtures is osmotic depletion: when the separation of the surfaces of two large spheres {(colloids)} is less than the diameter of the small ones {(polymers)}, the depletion of the latter from the gap between the colloids leads to an unbalanced osmotic pressure which, in turn, results in an effective attraction between the two large spheres. In this way, a fluid phase rich in large spheres may form  and coexist with another fluid phase rich in the small ones.
If the mixture is a nonadditive hard-sphere {(NAHS)} one with positive nonadditivity {(}which means that the distance of closest approach between two spheres of different species  is larger than the sum of their radii{),} it is well known that for sufficiently large nonadditivity it will present fluid-fluid separation into two fluid phases of different composition. Thus, one may affirm that fluid-fluid phase separation in HS mixtures may be due either to size asymmetry or to positive nonadditivity or to a combination of both effects.\cite{BH97}

The Asakura--Oosawa (AO) model\cite{AO54,AO58} (also developed later independently by Vrij\cite{V76}) was perhaps the first work in which the depletion mechanism was described. This model was introduced to study mixtures of colloidal particles and nonadsorbing polymers {(see Ref.\ \onlinecite{BVS14} for a recent review and references therein)}. In it, the colloid-colloid interactions were taken to be those of HS, the polymer-polymer interactions were assumed to vanish (i.e., the polymers were considered as an ideal gas of point particles), and the polymer-colloid interactions were of the excluded volume type and took into account the radius of gyration of the polymer. This model may also be considered as a limiting case of a NAHS mixture with positive nonadditivity. Hence, the AO model  incorporates simultaneously the two mechanisms responsible for fluid-fluid phase separation in HS systems, namely size asymmetry (leading to osmotic depletion) and positive nonadditivity. Not surprisingly, there is already a wealth of studies in the literature concerning the phase behavior of polymer-colloid mixtures using the AO model.
On the theoretical side, mention must be made of the thermodynamic perturbation theory approach,\cite{GHR83,LFH00,VPR05} of the free-volume {(FV)} theory,\cite{LPPSW92,IOPP95,FS05a,FS05b} and of the density functional approach.\cite{SLBE00,SLBE02,HS10} On the other hand, another  rather successful and often employed approach in the colloids literature is that of coarse graining.  {T}he idea is to integrate out the irrelevant degrees of freedom ({those associated with} the small particles) to end up with a one-component system of colloidal particles described by effective interactions (see for instance Refs.\ \onlinecite{DBE99}, \onlinecite{DRE00}, \onlinecite{BES03} and references therein).
{An important} result in this context is the proof that for a size ratio $q$ (see below for its definition) equal to or smaller than $2/\sqrt{3}-1\simeq 0.1547$ the mapping to the effective one-component system is exact.\cite{DBE99,DRE00} {This threshold value has a geometric origin: if $q<2/\sqrt{3}-1$ a polymer can fit into the inner volume created by three colloids at contact.} Finally, simulations have also been performed using either the effective one-component depletion potential or the full AO binary mixture.\cite{MF91,MF94,BBF96,BLH02,DLL02,VH04a,VH04b,VHB05,VVPR06,FSD08,RE08,ZVBHV09,AWRE11,AP11,AW14,RGPZ14}

Our motivation to carry out the present study rests, however, on different grounds. Three of us have considered fluid-fluid demixing in binary AHS mixtures using the available information on the virial coefficients of those mixtures.\cite{HT04,HML10,HTS13} A similar approach was followed by Vlasov and Masters.\cite{VM03} Although non conclusive due to the small number of available virial coefficients,  these studies suggest that in AHS mixtures there {would} be no {(}stable {or metastable)} fluid-fluid phase separation. But such a conclusion is not free from controversy since it is not clear whether the limited knowledge of the first few virial coefficients allows one to get a fair picture of fluid-fluid demixing caused by a depletion interaction. The question then arises {whether} in the AO model, where there is certainly fluid-fluid demixing, the approach based on the use of the available virial coefficients is at grips with the known results for the critical consolute point. The major aim of this paper is to address this question and assess how well the virial expansion performs with respect to the critical behavior of the AO model. To this end, it is important to stress that both the second and third virial coefficients of the AO model are known and exact. As far as the fourth one is concerned, to our knowledge it has not been reported so far {that} it also turns out to be exact,  its explicit expression  {being} provided  {below}. {In general, however, higher} order virial coefficients must necessarily be evaluated numerically and here we will provide values of the fifth virial coefficient of the AO model for selected values of the size ratio. With this limited information, we will consider the virial series truncated consecutively after the second, third, fourth, and fifth virial coefficient and compute the critical constants for a few size ratios. The results will be subsequently compared with those coming out of computer simulations.

The paper is organized as follows. In  {Sec.\ \ref{sec2}} we introduce the AO model and provide the virial coefficients up to the fifth, this latter only for a few size ratios. This is followed in Sec.\ \ref{sec3} by the computation of the critical constants and their representation in different thermodynamic planes. The paper is closed in  {Sec.\ \ref{sec4}} with some discussion of the results and a few concluding remarks.

\section{Virial coefficients of the Asakura--Oosawa Model}
\label{sec2}

Consider a binary fluid mixture of $N=N_1+ N_2$ {spheres} (colloids+polymers)  {in a}  volume $V$. In this mixture, the distance of closest approach between spheres of species $i$ and $j$, denoted by $\sigma_{ij}$, is such that $\sigma_{11}=\sigma_1$, $\sigma_{22}=0$, and $\sigma_{12}=\frac{1}{2}\sigma_1(1+q)$, with the size ratio $q$  acting as the (positive) nonadditivity parameter. This NAHS mixture  {defines} the well known AO model.\cite{BVS14}

The usual virial expansion of the compressibility factor of this mixture reads
\beq
Z\equiv\frac{p}{\rho k_B T}=1+\sum_{j=2}^{\infty}B_j(x_1,q) \rho^{j-1},
\label{comp}
\eeq
where $p$ is the pressure, $\rho$=$N/V$ is the number density, $k_B$ is the Boltzmann constant, $T$ is the absolute temperature, and it has been made explicit that the virial coefficients $B_j(x_1,q)$ {(in units of $\sigma_1^{3(j-1)}$)} depend only on {the mole fraction $x_1={N_1}/{N}$ of the colloids} and on the size ratio $q$. For later convenience, we further introduce at this stage  {the reduced pressure $p^*\equiv p\sigma_1^3/k_BT$,} the colloid packing fraction $\eta_c\equiv \frac{\pi}{6}\rho x_1 \sigma_1^3$, the (effective) polymer packing fraction $\eta_p\equiv \frac{\pi}{6}\rho x_2 \sigma_1^3 q^3$ {(where $x_2=1-x_1$ is the polymer mole fraction)}, and the (effective) \emph{reservoir} polymer packing fraction $\eta_p^r\equiv e^{\mu_2{/k_BT}}\frac{\pi}{6} {\sigma_1^3 q^3}/{{\Lambda_2^3}}{=e^{\mu_2^\ex/k_BT}\eta_p}$, where  $\mu_2$ is the chemical potential of the polymer{s}, {$\mu_2^\ex$ is its excess part}, and ${\Lambda_i}$ is the thermal de Broglie wavelength of {species $i$}.

\subsection{{Second, third, and fourth virial coefficients}}

The second and third virial coefficients of a NAHS mixture with positive nonadditivity are exact and well known.\cite{H96} It is a simple matter to obtain the corresponding expressions for the {AO} model from them, namely
\beq
B_2{(x_1,q)}=x_1^2 B_{11}+2x_1x_2B_{12}{(q)},
\label{B2}
\eeq
\beq
B_3{(x_1,q)}=x_1^3C_{111}+3x_1^2x_2C_{112}{(q)},
\label{B3}
\eeq
with {the composition-independent coefficients} (in units of $\sigma_1^3$)
\beq
B_{11}=\frac{\pi}{6}{4},
\eeq
\beq
B_{12}=\frac{\pi}{6}\frac{(1+q)^3}{2},
\eeq
and (in units of $\sigma_1^6$)
\beq
C_{111}= \left(\frac{\pi}{6}\right)^2{10},
\eeq
\beq
C_{112}=\left(\frac{\pi}{6}\right)^2\frac{1+6q+15q^2+8q^3}{3}.
\eeq
{For a general binary mixture, additional terms $x_2^2 B_{22}$ and $3x_1x_2^2C_{122}+x_2^3C_{222}$ should be included in Eqs.\ \eqref{B2} and \eqref{B3}, respectively. On the other hand, thanks to the property $\sigma_{22}=0$, the composition-independent coefficients $B_{22}$, $C_{122}$, and $C_{222}$ vanish in the AO model.}

The fourth virial coefficient for a \emph{general} NAHS mixture is not known exactly.
{However, in the special case of the AO model the fourth virial coefficient reads}
\beq
B_4{(x_1,q)}=x_1^4D_{1111}+4x_1^3x_2D_{1112}{(q)}+6x_1^2x_2^2D_{1122}{(q)},
\eeq
{since, analogously to  what happens in Eqs.\ \eqref{B2} and \eqref{B3},  the missing coefficients $D_{1222}$ and $D_{2222}$ vanish due to $\sigma_{22}=0$}. $D_{1111}$ corresponds to the fourth virial  coefficient of the {one-component} HS fluid, whose analytical  {value} (in units of $\sigma_1^9$) reads
\beq
D_{1111}=\left(\frac{\pi}{6}\right)^3 b_4,
\eeq
where
\beq
b_4=\frac{219\sqrt{2}-712\pi+4131\tan^{-1}\sqrt{2}}{35\pi}\simeq 18.3648.
\eeq
In terms of   Mayer diagrams, the coefficients $D_{1112}$  and $D_{1122}$ {in the AO model} are given by
\tikzstyle{fc} = [circle, minimum width=5pt, fill, inner sep=0pt]
\tikzstyle{ec} = [circle, minimum width=0pt, draw, inner sep=0pt]
%
\def\asquare{
	\begin{scope}
	\node[ec] (tr) at (45:1cm) {};
	\node[fc] (tl) at (135:1cm) {};
	\node[ec] (bl) at (225:1cm) {};
	\node[fc] (br) at (315:1cm) {};
	\end{scope}
}
\def\bsquare{
	\begin{scope}
	\node[fc] (tr) at (45:1cm) {};
	\node[fc] (tl) at (135:1cm) {};
	\node[fc] (bl) at (225:1cm) {};
	\node[ec] (br) at (315:1cm) {};
	\end{scope}
}
\def\apentagon{
	\begin{scope}
	\node[fc] (mr) at (18:1cm) {};
	\node[ec] (tt) at (90:1cm) {};
	\node[fc] (ml) at (162:1cm) {};
	\node[ec] (bl) at (234:1cm) {};
	\node[fc] (br) at (306:1cm) {};
	\end{scope}
}
\def\bpentagon{
	\begin{scope}
	\node[fc] (mr) at (18:1cm) {};
	\node[fc] (tt) at (90:1cm) {};
	\node[ec] (ml) at (162:1cm) {};
	\node[fc] (bl) at (234:1cm) {};
	\node[ec] (br) at (306:1cm) {};
	\end{scope}
}
\def\cpentagon{
	\begin{scope}
	\node[ec] (mr) at (18:1cm) {};
	\node[fc] (tt) at (90:1cm) {};
	\node[ec] (ml) at (162:1cm) {};
	\node[fc] (bl) at (234:1cm) {};
	\node[fc] (br) at (306:1cm) {};
	\end{scope}
}

\def\dpentagon{
	\begin{scope}
	\node[fc] (mr) at (18:1cm) {};
	\node[ec] (tt) at (90:1cm) {};
	\node[fc] (ml) at (162:1cm) {};
	\node[ec] (bl) at (234:1cm) {};
	\node[ec] (br) at (306:1cm) {};
	\end{scope}
}

\def\Apentagon{
	\begin{scope}
	\node[fc] (mr) at (18:1cm) {};
	\node[fc] (tt) at (90:1cm) {};
	\node[ec] (ml) at (162:1cm) {};
	\node[fc] (bl) at (234:1cm) {};
	\node[fc] (br) at (306:1cm) {};
	\end{scope}
}
\def\Bpentagon{
	\begin{scope}
	\node[fc] (mr) at (18:1cm) {};
	\node[ec] (tt) at (90:1cm) {};
	\node[fc] (ml) at (162:1cm) {};
	\node[fc] (bl) at (234:1cm) {};
	\node[fc] (br) at (306:1cm) {};
	\end{scope}
}
\def\Cpentagon{
	\begin{scope}
	\node[fc] (mr) at (18:1cm) {};
	\node[fc] (tt) at (90:1cm) {};
	\node[fc] (ml) at (162:1cm) {};
	\node[ec] (bl) at (234:1cm) {};
	\node[fc] (br) at (306:1cm) {};
	\end{scope}
}
\def\Dpentagon{
	\begin{scope}
	\node[fc] (mr) at (18:1cm) {};
	\node[fc] (tt) at (90:1cm) {};
	\node[fc] (ml) at (162:1cm) {};
	\node[fc] (bl) at (234:1cm) {};
	\node[ec] (br) at (306:1cm) {};
	\end{scope}
}
%
\def\paa#1{
\begin{scope}[shift={#1}]
	\apentagon
	\draw (br) -- (mr) -- (tt) -- (ml) -- (bl) -- (br);
\end{scope}
}
\def\pab#1{
\begin{scope}[shift={#1}]
	\apentagon
	\draw (br) -- (mr) -- (tt) -- (ml) -- (bl) -- (br);
	\draw (ml) -- (mr);
\end{scope}
}
\def\pac#1{
\begin{scope}[shift={#1}]
	\apentagon
	\draw (br) -- (mr) -- (tt) -- (ml) -- (bl) -- (br);
	\draw (ml) -- (br);
	\draw (bl) -- (mr);
\end{scope}
}
\def\pad#1{
\begin{scope}[shift={#1}]
	\apentagon
	\draw (br) -- (mr) -- (tt) -- (ml) -- (bl) -- (br);
	\draw (ml) -- (br);
	\draw (bl) -- (mr);
	\draw (ml) -- (mr);
\end{scope}
}
\def\pae#1{
\begin{scope}[shift={#1}]
	\apentagon
	\draw (br) -- (mr) -- (tt) -- (ml) -- (bl);
	\draw (ml) -- (br);
	\draw (bl) -- (mr);
\end{scope}
}
\def\paf#1{
\begin{scope}[shift={#1}]
	\apentagon
	\draw (br) -- (mr) -- (tt) -- (ml) -- (bl);
	\draw (ml) -- (br);
	\draw (bl) -- (mr);
	\draw (ml) -- (mr);
\end{scope}
}
%
\def\pba#1{
\begin{scope}[shift={#1}]
	\bpentagon
	\draw (br) -- (mr) -- (tt) -- (ml) -- (bl) -- (br);
	\draw (ml) -- (mr);
\end{scope}
}
\def\pbb#1{
\begin{scope}[shift={#1}]
	\bpentagon
	\draw (br) -- (mr) -- (tt) -- (ml) -- (bl) -- (br);
	\draw (tt) -- (bl);
	\draw (tt) -- (br);
\end{scope}
}
\def\pbc#1{
\begin{scope}[shift={#1}]
	\bpentagon
	\draw (br) -- (mr) -- (tt) -- (ml) -- (bl) -- (br);
	\draw (tt) -- (bl);
	\draw (tt) -- (br);
	\draw (ml) -- (mr);
\end{scope}
}
\def\pca#1{
\begin{scope}[shift={#1}]
	\cpentagon
	\draw (br) -- (mr) -- (tt) -- (ml) -- (bl) -- (br);
	\draw (tt) -- (bl);
	\draw (tt) -- (br);
\end{scope}
}
\def\pcb#1{
\begin{scope}[shift={#1}]
	\cpentagon
	\draw (br) -- (mr) -- (tt) -- (ml) -- (bl) -- (br);
	\draw (ml) -- (br);
	\draw (mr) -- (bl);
\end{scope}
}
\def\pcc#1{
\begin{scope}[shift={#1}]
	\cpentagon
	\draw (br) -- (mr) -- (tt) -- (ml) -- (bl) -- (br);
	\draw (ml) -- (br);
	\draw (mr) -- (bl);
	\draw (tt) -- (bl);
	\draw (tt) -- (br);
\end{scope}
}
\def\pcd#1{
\begin{scope}[shift={#1}]
	\cpentagon
	\draw (br) -- (mr) -- (tt) -- (ml) -- (bl);
	\draw (ml) -- (br);
	\draw (mr) -- (bl);
\end{scope}
}
\def\pAa#1{
\begin{scope}[shift={#1}]
	\Apentagon
	\draw (br) -- (mr) -- (tt) -- (ml) -- (bl) -- (br);
\end{scope}
}
%
\def\pAb#1{
\begin{scope}[shift={#1}]
	\Apentagon
	\draw (br) -- (mr) -- (tt) -- (ml) -- (bl) -- (br);
	\draw (ml) -- (mr);
\end{scope}
}
%
\def\pAc#1{
\begin{scope}[shift={#1}]
	\Apentagon
	\draw (br) -- (mr) -- (tt) -- (ml) -- (bl) -- (br);
	\draw (bl) -- (tt);
	\draw (br) -- (tt);
\end{scope}
}
%
\def\pAd#1{
\begin{scope}[shift={#1}]
	\Apentagon
	\draw (br) -- (mr) -- (tt) -- (ml) -- (bl) -- (br);
	\draw (bl) -- (mr);
	\draw (br) -- (ml);
\end{scope}
}
%
\def\pAe#1{
\begin{scope}[shift={#1}]
	\Apentagon
	\draw (br) -- (mr) -- (tt) -- (ml) -- (bl) -- (br);
	\draw (ml) -- (mr);
	\draw (bl) -- (mr);
	\draw (br) -- (ml);
\end{scope}
}
%
\def\pAf#1{
\begin{scope}[shift={#1}]
	\Apentagon
	\draw (br) -- (mr) -- (tt) -- (ml) -- (bl) -- (br);
	\draw (ml) -- (mr);
	\draw (bl) -- (tt);
	\draw (br) -- (tt);
\end{scope}
}
%
\def\pAg#1{
\begin{scope}[shift={#1}]
	\Apentagon
	\draw (br) -- (mr) -- (tt) -- (ml) -- (bl) -- (br);
	\draw (bl) -- (tt);
	\draw (br) -- (tt);
	\draw (bl) -- (mr);
	\draw (br) -- (ml);
\end{scope}
}
%
\def\pAh#1{
\begin{scope}[shift={#1}]
	\Apentagon
	\draw (br) -- (mr) -- (tt) -- (ml) -- (bl);
	\draw (bl) -- (mr);
	\draw (br) -- (ml);
\end{scope}
}
%
\def\pAi#1{
\begin{scope}[shift={#1}]
	\Apentagon
	\draw (br) -- (mr) -- (tt) -- (ml) -- (bl);
	\draw (ml) -- (mr);
	\draw (bl) -- (mr);
	\draw (br) -- (ml);
\end{scope}
}
%
%
\def\pBa#1{
\begin{scope}[shift={#1}]
	\Bpentagon
	\draw (br) -- (mr) -- (tt) -- (ml) -- (bl) -- (br);
	\draw (ml) -- (mr);
\end{scope}
}
%
\def\pBb#1{
\begin{scope}[shift={#1}]
	\Bpentagon
	\draw (br) -- (mr) -- (tt) -- (ml) -- (bl) -- (br);
	\draw (bl) -- (tt);
	\draw (br) -- (tt);
\end{scope}
}
%
\def\pBc#1{
\begin{scope}[shift={#1}]
	\Bpentagon
	\draw (br) -- (mr) -- (tt) -- (ml) -- (bl) -- (br);
	\draw (bl) -- (mr);
	\draw (br) -- (ml);
\end{scope}
}
%
\def\pBd#1{
\begin{scope}[shift={#1}]
	\Bpentagon
	\draw (br) -- (mr) -- (tt) -- (ml) -- (bl) -- (br);
	\draw (ml) -- (mr);
	\draw (bl) -- (mr);
	\draw (br) -- (ml);
\end{scope}
}
%
\def\pBe#1{
\begin{scope}[shift={#1}]
	\Bpentagon
	\draw (br) -- (mr) -- (tt) -- (ml) -- (bl) -- (br);
	\draw (ml) -- (mr);
	\draw (bl) -- (tt);
	\draw (br) -- (tt);
\end{scope}
}
%
\def\pBf#1{
\begin{scope}[shift={#1}]
	\Bpentagon
	\draw (br) -- (mr) -- (tt) -- (ml) -- (bl) -- (br);
	\draw (bl) -- (tt);
	\draw (br) -- (tt);
	\draw (bl) -- (mr);
	\draw (br) -- (ml);
\end{scope}
}
%
\def\pBg#1{
\begin{scope}[shift={#1}]
	\Bpentagon
	\draw (br) -- (mr) -- (tt) -- (ml) -- (bl) -- (br);
	\draw (ml) -- (mr);
	\draw (bl) -- (tt);
	\draw (br) -- (tt);
	\draw (bl) -- (mr);
	\draw (br) -- (ml);
\end{scope}
}
%
\def\pBh#1{
\begin{scope}[shift={#1}]
	\Bpentagon
	\draw (br) -- (mr) -- (tt) -- (ml) -- (bl);
	\draw (bl) -- (mr);
	\draw (br) -- (ml);
\end{scope}
}
%
\def\pBi#1{
\begin{scope}[shift={#1}]
	\Bpentagon
	\draw (br) -- (mr) -- (tt) -- (ml) -- (bl);
	\draw (ml) -- (mr);
	\draw (bl) -- (mr);
	\draw (br) -- (ml);
\end{scope}
}
%
%
\def\pCa#1{
\begin{scope}[shift={#1}]
	\Cpentagon
	\draw (br) -- (mr) -- (tt) -- (ml) -- (bl) -- (br);
	\draw (ml) -- (mr);
\end{scope}
}
%
\def\pCb#1{
\begin{scope}[shift={#1}]
	\Cpentagon
	\draw (br) -- (mr) -- (tt) -- (ml) -- (bl) -- (br);
	\draw (bl) -- (tt);
	\draw (br) -- (tt);
\end{scope}
}
%
\def\pCc#1{
\begin{scope}[shift={#1}]
	\Cpentagon
	\draw (br) -- (mr) -- (tt) -- (ml) -- (bl) -- (br);
	\draw (bl) -- (mr);
	\draw (br) -- (ml);
\end{scope}
}
%
\def\pCd#1{
\begin{scope}[shift={#1}]
	\Cpentagon
	\draw (br) -- (mr) -- (tt) -- (ml) -- (bl) -- (br);
	\draw (ml) -- (mr);
	\draw (bl) -- (tt);
	\draw (br) -- (tt);
\end{scope}
}
%
\def\pCe#1{
\begin{scope}[shift={#1}]
	\Cpentagon
	\draw (br) -- (mr) -- (tt) -- (ml) -- (bl) -- (br);
	\draw (bl) -- (tt);
	\draw (br) -- (tt);
	\draw (bl) -- (mr);
	\draw (br) -- (ml);
\end{scope}
}
%
\def\pCf#1{
\begin{scope}[shift={#1}]
	\Cpentagon
	\draw (br) -- (mr) -- (tt) -- (ml) -- (bl) -- (br);
	\draw (ml) -- (mr);
	\draw (bl) -- (mr);
	\draw (br) -- (ml);
\end{scope}
}
%

%
\def\pDa#1{
\begin{scope}[shift={#1}]
	\Dpentagon
	\draw (br) -- (mr) -- (tt) -- (ml) -- (bl) -- (br);
	\draw (ml) -- (mr);
	\draw (bl) -- (mr);
	\draw (br) -- (ml);
\end{scope}
}
%
\def\sa#1{
\begin{scope}[shift={#1}]
	\asquare
	\draw (br) -- (tr) -- (tl) -- (bl) -- (br);
\end{scope}
}
\def\sc#1{
\begin{scope}[shift={#1}]
	\asquare
	\draw (br) -- (tr) -- (tl) -- (bl) -- (br);
	\draw (tl) -- (br);
\end{scope}
}
%
%
\def\sb#1{
\begin{scope}[shift={#1}]
	\asquare
	\draw (br) -- (tr) -- (tl) -- (bl) -- (br);
\end{scope}
}
\def\sba#1{
\begin{scope}[shift={#1}]
	\bsquare
	\draw (br) -- (tr) -- (tl) -- (bl) -- (br);
\end{scope}
}
\def\sbb#1{
\begin{scope}[shift={#1}]
	\bsquare
	\draw (br) -- (tr) -- (tl) -- (bl) -- (br);
	\draw (bl) -- (tr);
\end{scope}
}
\def\sbc#1{
\begin{scope}[shift={#1}]
	\bsquare
	\draw (br) -- (tr) -- (tl) -- (bl) -- (br);
	\draw (br) -- (tl);
\end{scope}
}
\def\sbd#1{
\begin{scope}[shift={#1}]
	\bsquare
	\draw (br) -- (tr) -- (tl) -- (bl) -- (br);
	\draw (br) -- (tl);
	\draw (bl) -- (tr);
\end{scope}
}

\def\pcdnew#1{
\begin{scope}[shift={#1}]
	\dpentagon
	\draw (br) -- (mr) -- (tt) -- (ml) -- (bl);
	\draw (ml) -- (br);
	\draw (mr) -- (bl);
\end{scope}
}
\def\pcdnewtwo#1{
\begin{scope}[shift={#1}]
	\dpentagon
	\draw (br) -- (mr) -- (tt) -- (ml) -- (bl);
	\draw (ml) -- (mr);
	\draw (bl) -- (mr);
	\draw (br) -- (ml);
\end{scope}
}

\begin{equation}
D_{1112} = - \frac{1}{8} \left(3\;
\begin{tikzpicture}[style=thick,scale=0.45, baseline] \sba{(0,0)} \end{tikzpicture}
+3 \;
\begin{tikzpicture}[style=thick,scale=0.45, baseline] \sbb{(0,0)} \end{tikzpicture}
+3 \;
\begin{tikzpicture}[style=thick,scale=0.45, baseline] \sbc{(0,0)} \end{tikzpicture}
+ \;
\begin{tikzpicture}[style=thick,scale=0.45, baseline] \sbd{(0,0)} \end{tikzpicture}\right)
,
\label{d1112}
\end{equation}
\begin{equation}
D_{1122} = - \frac{1}{8}\left(
\begin{tikzpicture}[style=thick,scale=0.45, baseline] \sa{(0,0)} \end{tikzpicture}
+\;
\begin{tikzpicture}[style=thick,scale=0.45, baseline] \sc{(0,0)} \end{tikzpicture}\right) .
\label{d1122}
\end{equation}
{Here and below, a circle represents a colloidal particle and a single vertex represents a polymer particle. For example,}
\bal
{\begin{tikzpicture}[style=thick,scale=0.45, baseline] \sba{(0,0)} \end{tikzpicture}=}&{\frac{1}{V}\int d\mathbf{r}_1\int d\mathbf{r}_2\int d\mathbf{r}_3\int d\mathbf{r}_4\, f_{11}(r_{12})f_{11}(r_{13})}\nn
&{\times f_{12}(r_{24})f_{12}(r_{34}),}
\eal
{where $f_{ij}(r)=-\Theta(\sigma_{ij}-r)$ is the Mayer function corresponding to the interaction $ij$, $\Theta(x)$ being the Heaviside step function.}

Interestingly enough, the Mayer diagrams {in Eqs.\ \eqref{d1112} and \eqref{d1122}} are  {equivalent} to the corresponding ones appearing in general AHS mixtures {since they do not actually depend on the value of $\sigma_{22}$}. In fact, $D_{1112}$ is exactly the same {as for general mixtures} and it turns out that there exist analytical {expressions}\cite{B98,LK09,U11} for this partial coefficient, {as well as}  for {the two}  diagrams appearing in $D_{1122}$ as given by Eq.\ (\ref{d1122}).   Taking into account those results, {one finds that $D_{1122}$ and $D_{1112}$ are} given (in units of $\sigma_1^9$) by
\beq
D_{1122}=-\left(\frac{\pi}{6}\right)^3{q^5}\left(\frac{27}{20}+\frac{12q}{5}
+\frac{51q^2}{35}+\frac{51q^3}{140}+\frac{17q^4}{420}\right),
\label{D1122}
\eeq
\beq
D_{1112}=\begin{cases}
D_{1112}^{(1)},&q\leq 2/\sqrt{3}-1,\\
D_{1112}^{(1)}+D_{1112}^{(2)},&q>2/\sqrt{3}-1,
\end{cases}
\eeq
\bal
D_{1112}^{(1)}=&\left(\frac{\pi}{6}\right)^3\left(\frac{1}{4}+\frac{9q}{4}+9q^2
+\frac{21q^3}{4}+\frac{27q^4}{8}+\frac{27q^5}{40}\right.
\nn
&
\left.-\frac{27q^6}{5}-\frac{162q^7}{35}-\frac{81q^8}{56}-\frac{9q^9}{56}\right),
\label{6a}
\eal
\bal
D_{1112}^{(2)}=&\left(\frac{\pi}{6}\right)^3\frac{1}{280\pi}\Big[\frac{Q}{12}\left(10Q^6-
51Q^4+210Q^2+6976\right)
\nn
&
-486P_1(Q^2+9)+\frac{q+1}{3} P_2\left(5Q^8-28Q^6\right.\nn
& \left.+129Q^4-124Q^2
+11378\right)
\Big],
\label{DeltaB4}
\eal
where $Q\equiv\sqrt{3q^2+6q-1}$, $P_1\equiv \tan^{-1} Q$, and $P_2\equiv \tan^{-1}\left[Q/(q+1)\right]$.
Therefore, we now have the exact analytical results up to the fourth virial coefficient of the AO model.

\subsection{{Fifth virial coefficient}}
Concerning the fifth virial coefficient,  the terms that survive in the AO model are
\bal
B_5{(x_1,q)}=&x_1^5E_{11111}+5x_1^4x_2E_{11112}{(q)}+10x_1^3x_2^2 E_{11122}{(q)}\nn
&
{+10x_1^2x_2^3 E_{11222}{(q)}}.
\eal
The condition $\sigma_{22}=0$ implies that the coefficients {$E_{12222}$ and $E_{22222}$}  vanish.
The partial term $E_{11111}$ corresponds to the fifth virial coefficient of the {one-component} HS fluid whose numerical value (in units of $\sigma_1^{12}$) is
\beq
E_{11111}=\left(\frac{\pi}{6}\right)^4b_5,\quad b_5\simeq 28.224512.
\eeq
{As for {$E_{11112}$, $E_{11122}$, and $E_{11222}$}, they can be expressed in terms of Mayer diagrams as}
\begin{widetext}
\bal
E_{11112} =& - \frac{1}{30}\left(12
 \;
\begin{tikzpicture}[style=thick,scale=0.45, baseline] \pAa{(0,0)} \end{tikzpicture}
+12 \;
\begin{tikzpicture}[style=thick,scale=0.45, baseline] \pBa{(0,0)} \end{tikzpicture}
+24 \;
\begin{tikzpicture}[style=thick,scale=0.45, baseline] \pAb{(0,0)} \end{tikzpicture}
+24 \;
\begin{tikzpicture}[style=thick,scale=0.45, baseline] \pCa{(0,0)} \end{tikzpicture}
+12\;
\begin{tikzpicture}[style=thick,scale=0.45, baseline] \pBb{(0,0)} \end{tikzpicture}
+24 \;
\begin{tikzpicture}[style=thick,scale=0.45, baseline] \pAc{(0,0)} \end{tikzpicture}
+24 \;
\begin{tikzpicture}[style=thick,scale=0.45, baseline] \pCb{(0,0)} \end{tikzpicture}
\right.
\nn
&+6 \;
\begin{tikzpicture}[style=thick,scale=0.45, baseline] \pBc{(0,0)} \end{tikzpicture}
+12 \;
\begin{tikzpicture}[style=thick,scale=0.45, baseline] \pAd{(0,0)} \end{tikzpicture}
+12 \;
\begin{tikzpicture}[style=thick,scale=0.45, baseline] \pCc{(0,0)} \end{tikzpicture}
+6\;
\begin{tikzpicture}[style=thick,scale=0.45, baseline] \pBd{(0,0)} \end{tikzpicture}
+12 \;
\begin{tikzpicture}[style=thick,scale=0.45, baseline] \pAe{(0,0)} \end{tikzpicture}
+12 \;
\begin{tikzpicture}[style=thick,scale=0.45, baseline] \pCf{(0,0)} \end{tikzpicture}
+3\;
\begin{tikzpicture}[style=thick,scale=0.45, baseline] \pBe{(0,0)} \end{tikzpicture}
+12 \;
\begin{tikzpicture}[style=thick,scale=0.45, baseline] \pAf{(0,0)} \end{tikzpicture}
\nn
&
\left.
+4 \;
\begin{tikzpicture}[style=thick,scale=0.45, baseline] \pAg{(0,0)} \end{tikzpicture}
+6 \;
\begin{tikzpicture}[style=thick,scale=0.45, baseline] \pCe{(0,0)} \end{tikzpicture}
+ \;
\begin{tikzpicture}[style=thick,scale=0.45, baseline] \pBg{(0,0)} \end{tikzpicture}
+6\;
\begin{tikzpicture}[style=thick,scale=0.45, baseline] \pBh{(0,0)} \end{tikzpicture}
+4 \;
\begin{tikzpicture}[style=thick,scale=0.45, baseline] \pAh{(0,0)} \end{tikzpicture}
+6 \;
\begin{tikzpicture}[style=thick,scale=0.45, baseline] \pBi{(0,0)} \end{tikzpicture}
+4\;
\begin{tikzpicture}[style=thick,scale=0.45, baseline] \pAi{(0,0)} \end{tikzpicture}\right),
\eal
\bal
E_{11122} =& -\frac{1}{30}\left(6 \;
\begin{tikzpicture}[style=thick,scale=0.45, baseline] \paa{(0,0)} \end{tikzpicture}
+12 \;
\begin{tikzpicture}[style=thick,scale=0.45, baseline] \pab{(0,0)} \end{tikzpicture}
+6 \;
\begin{tikzpicture}[style=thick,scale=0.45, baseline] \pac{(0,0)} \end{tikzpicture}
+6 \;
\begin{tikzpicture}[style=thick,scale=0.45, baseline] \pad{(0,0)} \end{tikzpicture}
+3\;
\begin{tikzpicture}[style=thick,scale=0.45, baseline] \pae{(0,0)} \end{tikzpicture}
+3 \;
\begin{tikzpicture}[style=thick,scale=0.45, baseline] \paf{(0,0)} \end{tikzpicture}
+12 \;
\begin{tikzpicture}[style=thick,scale=0.45, baseline] \pba{(0,0)} \end{tikzpicture}
+12 \;
\begin{tikzpicture}[style=thick,scale=0.45, baseline] \pbb{(0,0)} \end{tikzpicture}\right.
\nn
&\left.
+3\;
\begin{tikzpicture}[style=thick,scale=0.45, baseline] \pbc{(0,0)} \end{tikzpicture}
+6\;
\begin{tikzpicture}[style=thick,scale=0.45, baseline] \pca{(0,0)} \end{tikzpicture}
+3 \;
\begin{tikzpicture}[style=thick,scale=0.45, baseline] \pcb{(0,0)} \end{tikzpicture}
+ \;
\begin{tikzpicture}[style=thick,scale=0.45, baseline] \pcc{(0,0)} \end{tikzpicture}
+\;
\begin{tikzpicture}[style=thick,scale=0.45, baseline] \pcd{(0,0)} \end{tikzpicture}\right) ,
\eal
\end{widetext}
\beq
{E_{11222}=-\frac{1}{30}\left(
\begin{tikzpicture}[style=thick,scale=0.45, baseline] \pcdnew{(0,0)} \end{tikzpicture}
+\;
\begin{tikzpicture}[style=thick,scale=0.45, baseline] \pcdnewtwo{(0,0)} \end{tikzpicture}
\right).}
\eeq
{As happened with $D_{1112}$, t}he number of irreducible graphs of $E_{11112}$ remains unchanged with respect to what happens in the general binary mixture. {On the other hand, similarly to the case of $D_{1122}$,}   the number of graphs representing $E_{11122}$ {and $E_{11222}$} is greatly reduced with respect to the complete set.
{Exploiting the fact that only the effective colloid-colloid pair potential\cite{DBE99} contributes to the osmotic pressure to second order in the colloid density, it is possible to prove that}
\bal
{E_{11222}=}&{-\left(\frac{\pi}{6}\right)^4 \frac{q^7}{8400}\left(3240+7695q+6780q^2+2706q^3\right.}\nn
&{\left.+492q^4+41q^5\right)}.
\label{E11222}
\eal

{To} our knowledge there are no analytical results for th{e {coefficients} $E_{11112}$ and $E_{11122}$}.
Therefore, {those} coefficients have to be computed numerically for each $q$.
In order to do so, {a} standard Monte Carlo (MC) numerical integration procedure, similar to those of Refs.\ \onlinecite{SFG96} and \onlinecite{SFG98}, was employed.
The algorithm produces a significant set of configurations compatible with the Mayer graph one wants to evaluate.
We first fix particle $1$ {(by convention, a colloid)}  at the origin and sequentially deposit the remaining four particles {(colloids or polymers, depending on the graph)} at random but in such a way that particle $i+1$ overlaps with particle $i$ (where $i=1,2,3,4$). This procedure generates a ``trial configuration,'' that is, an open chain of overlapping particles. A ``successful configuration'' is a closed-chain configuration where particle $1$ overlaps with particle $5$ and the residual cross-linked ``bonds''  {that} are present in the Mayer graph that is being calculated are also retrieved.
The ratio of the number of successful configurations ($N_s$) to the total number of trial configurations ($N_t$) yields, asymptotically, the value of the cluster integral relative to that of the open-chain graph.  {The latter}, in turn, is trivially related to a product of the partial second-order virial coefficients $B_{\alpha \beta}$.

The numerical accuracy of the MC results obviously depends on the total number of trial configurations.
The {relative} error on the cluster integral $J$ is estimated as {$\sqrt{J(J-1)/N_t}$}.
However, as a result of the accumulation of statistically independent errors, the global uncertainty affecting the partial virial coefficients is higher than the error estimated for each cluster integral that enters {its} expression.
A typical MC run consisted of $4 \times 10^{11}$--$1 \times 10^{13}$ independent moves, depending on the {the value of $q$}.
In order to produce reliable pseudorandom numbers, we adopted the Mersenne Twister MT19937 pseudorandom number generator,\cite{MN98} which is characterized by a very long period ($2^{19937}-1$).

{The numerically computed values of $E_{11112}$ and $E_{11122}$ are presented in Table \ref{tab1} for particular values of the size ratio $q$. As a confidence test, we have numerically computed the partial {virial coefficients $D_{1112}$, $D_{1122}$, and $E_{11222}$} with the same MC method and for the same values of $q$ as in the cases of $E_{11112}$ and $E_{11122}$. Comparison with  exact expressions \eqref{D1122}--\eqref{DeltaB4} {and \eqref{E11222}} shows deviations smaller than the estimated error bars.}

\begin{table}
\caption{Numerical values of the partial coefficients $E_{11112}$ and $E_{11122}$ (in units of $\sigma_{1}^{12}$) for some values of the size ratio $q$. The error on the
last significant figure is enclosed in parentheses.
\label{tab1}}
\begin{ruledtabular}
\begin{tabular}{ccc}
 $q$ &  $E_{11112}$ & $E_{11122}$ \\
\hline
 $0.05$ &  $0.0267(6)$ &     $\approx -5 \times 10^{-8}$ \\
 $0.10$ &   $0.0437(4)$ &     $-0.0000043(9)$               \\
 $0.15$ &   $0.0666(8)$ &    $-0.000034(8)$                \\
 $0.40$ &   $0.296(4)$  &    $-0.0075(2)$                  \\
$0.56$ &   $0.575(5)$  &     $-0.0546(5)$                  \\
 $0.80$ &  $1.257(7)$  &     $-0.505(4)$                   \\
 $1.00$&$E_{11111}$&$-2.18(3)$ \\
\end{tabular}
\end{ruledtabular}
\end{table}

\section{The critical consolute point}
\label{sec3}

In order to study the critical behavior, we consider the Helmholtz free energy per particle {$f$}. If  virial expansion  (\ref{comp}) is truncated after $n$ terms, this quantity reads
\beq
\frac{f}{{k_BT}}=x_1\ln\left(x_1\rho{\Lambda_1^3}\right)+x_2\ln\left(x_2\rho{\Lambda_2^3}\right)-1+\sum_{j=2}^{n} \frac{B_j}{j-1}\rho^{j-1}.
\eeq
Considering partial derivatives of $f$ with respect to composition $x$ (where $x=x_1$) and specific volume $v$ (where $v=1/\rho$), the  conditions required for determining the critical point are
\beq
f_{xx}f_{vv}-f_{xv}^2=0,
\label{cond1}
\eeq
\beq
f_{xxx}-3f_{xxv}\frac{f_{xv}}{f_{vv}}+3f_{xvv}\left(\frac{f_{xv}}{f_{vv}}\right)^2
-f_{vvv}\left(\frac{f_{xv}}{f_{vv}}\right)^3=0,
\label{cond2}
\eeq
{and the chemical potentials are obtained as}
\beq
{\mu_1=\mu_2+f_x,\quad \mu_2=f-v f_v-x f_x.}
\eeq

Table \ref{tab2} contains the results for the critical constants  $x_{\text{cr}}$,  $\eta_{c,\text{cr}}$, {$\eta_{p,\text{cr}}^r$}, and $p_{\text{cr}}^*$ for particular values of the size ratio $q$ when the virial expansion is truncated taking $n=2$, $3$, $4$, and $5$, respectively, and Eqs.\ (\ref{cond1}) and (\ref{cond2}) are solved simultaneously.  {The critical value of the polymer packing fraction can easily be obtained from the relationship $\eta_{p,\text{cr}}=\eta_{c,\text{cr}}q^3(1-x_{\text{cr}})/x_{\text{cr}}$.}

\begin{table}
\caption{Critical constants $x_{\text{cr}}$, $\eta_{c,\text{cr}}$, {$\eta_{p,\text{cr}}^r$}, and $p_{\text{cr}}^*$ as
obtained from the truncation of the virial expansion after the $n$th
virial coefficient for different $q$-values.
\label{tab2}}
\begin{ruledtabular}
\begin{tabular}{cccccc}
 $q$ & $n$&$x_{\text{cr}}$&$\eta_{c,\text{cr}}$ &{$\eta_{p,\text{cr}}^r$}&$p_{\text{cr}}^*$\\
\hline
 $0.05$ & $2$ &$0.0303$ &$0.2720$&$0.0015$&$22.9654$ \\
& $3$&$0.0298$&$0.4291$&$0.0032$&$50.2366$\\
& $4$&$0.0261$&$0.5328$&$0.0060$&$93.6496$\\
&$5$&$0.0232$&$0.6035$&$0.0097$&$150.878$\\
&&&&&\\
$0.10$&$2$&$0.0356$&$0.2503$&$0.0094$&$18.1964$\\
&$3$&$0.0389$&$0.3903$&$0.0185$&$35.9421$\\
&$4$&$0.0372$&$0.4793$&$0.0313$&$60.9726$\\
&$5$&$0.0356$&$0.5385$&$0.0461$&${90.2842}$\\
&&&&&\\
$0.15$&$2$&$0.0415$&$0.2309$&$0.0256$&$14.6051$\\
&$3$&$0.0494$&$0.3550$&$0.0457$&$26.3274$\\
&$4$&$0.0503$&$0.4312$&$0.0711$&$41.2659$\\
&$5$&$0.0505$&${0.4812}$&$0.0983$&${57.2493}$\\
&&&&&\\
$0.40$&$2$&$0.0779$&$0.1599$&$0.1879$&$5.7026$\\
&$3$&$0.1165$&$0.2273$&$0.2423$&$7.4719$\\
&$4$&$0.1382$&$0.2691$&$0.3013$&$9.3443$\\
&$5$&${0.1546}$&${0.3013}$&${0.3577}$&${11.0738}$\\
&&&&&\\
$0.56$&$2$&$0.1060$&$0.1294$&$0.3132$&$3.4830$\\
&$3$&$0.1658$&$0.1763$&$0.3582$&$4.0599$\\
&$4$&$0.2035$&$0.2097$&$0.4195$&$4.7754$\\
&$5$&${0.2345}$&${0.2386}$&${0.4817}$&${5.4596}$\\
&&&&&\\
$0.80$&$2$&$0.1520$&$0.0967$&$0.4853$&$1.8668$\\
&$3$&$0.2378$&$0.1255$&$0.4994$&$1.9647$\\
&$4$&$0.2971$&$0.1521$&$0.5628$&$2.2185$\\
&$5$&${0.3518}$&${0.1786}$&${0.6346}$&${2.4837}$\\
&&&&&\\
$1.00$&$2$&$0.1910$&$0.0773$&$0.6071$&$1.2044$\\
&$3$&$0.2903$&$0.0972$&$0.5964$&$1.2115$\\
&$4$&$0.3638$&$0.1202$&$0.6659$&$1.3525$\\
&$5$&${0.4382}$&${0.1456}$&${0.7481}$&${1.5068}$\\
\end{tabular}
\end{ruledtabular}
\end{table}

\begin{figure}
\includegraphics[width=8cm]{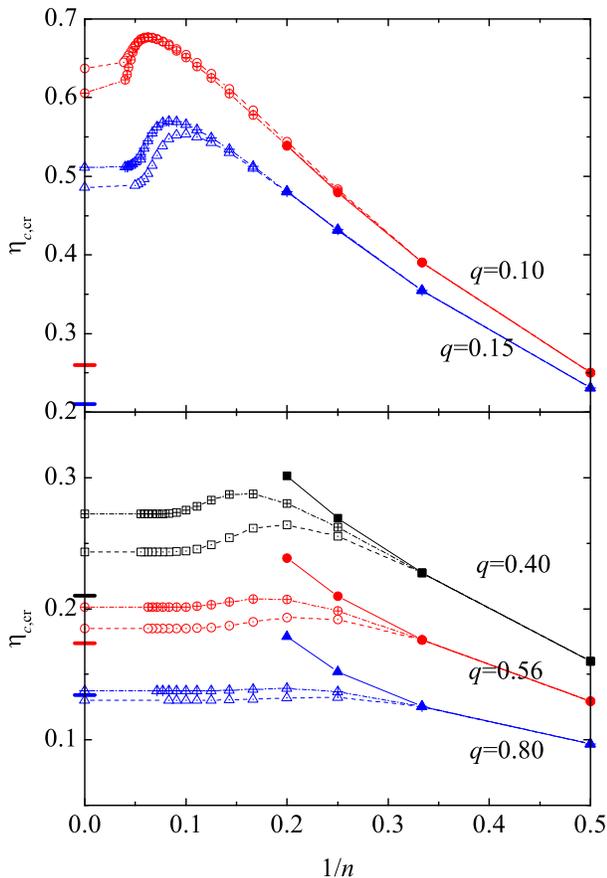}
\caption{Critical colloid packing fraction {($\eta_{c,\text{cr}}$)} as a function of the inverse of the number of retained virial coefficients in the virial series for different size ratios. {The solid symbols represent the values obtained from the exact second ($n=2$), third ($n=3$), and fourth ($n=4$),  as well as from our MC evaluation of the fifth ($n=5$) virial coefficient. The open and crossed symbols represent the values obtained by truncating the FV\cite{LPPSW92} and SHY\cite{SHY05,SHY10} analytical equations of state, respectively. The critical values provided by the full equations of state are represented at $1/n=0$ {and joined to the last truncated value by dashed and dash-dotted lines}. Finally, the short horizontal lines at $1/n=0$ represent the critical values obtained from MC simulations.\cite{VH04a,VH04b,VHB05,VVPR06,FSD08,ZVBHV09,AWRE11}}}
\label{fig1}
\end{figure}

\begin{figure}
\includegraphics[width=8cm]{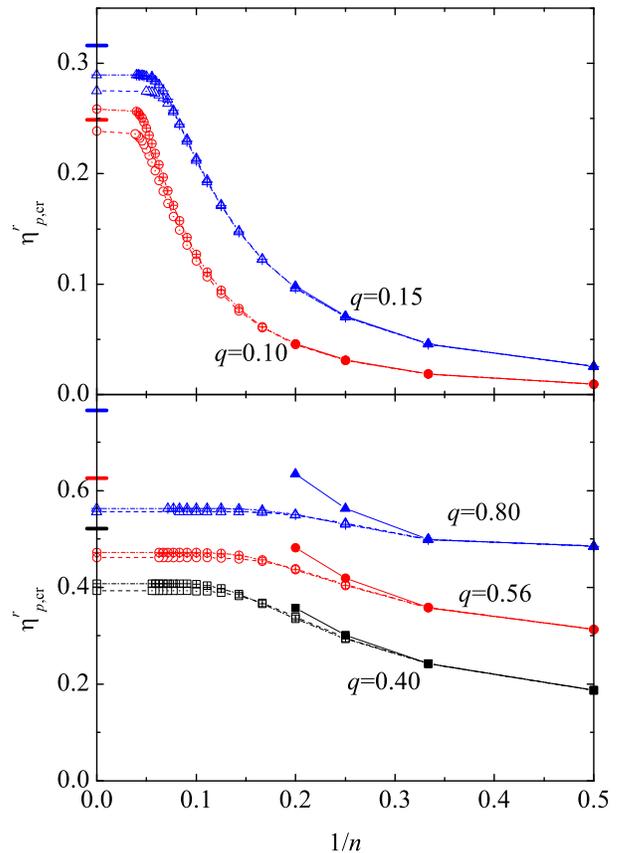}
\caption{{Same as in Fig.\ \protect\ref{fig1}, but for the} critical (effective) reservoir polymer packing fraction {($\eta_{p,\text{cr}}^r$)}.}
\label{fig2}
\end{figure}

\begin{figure}
\includegraphics[width=8cm]{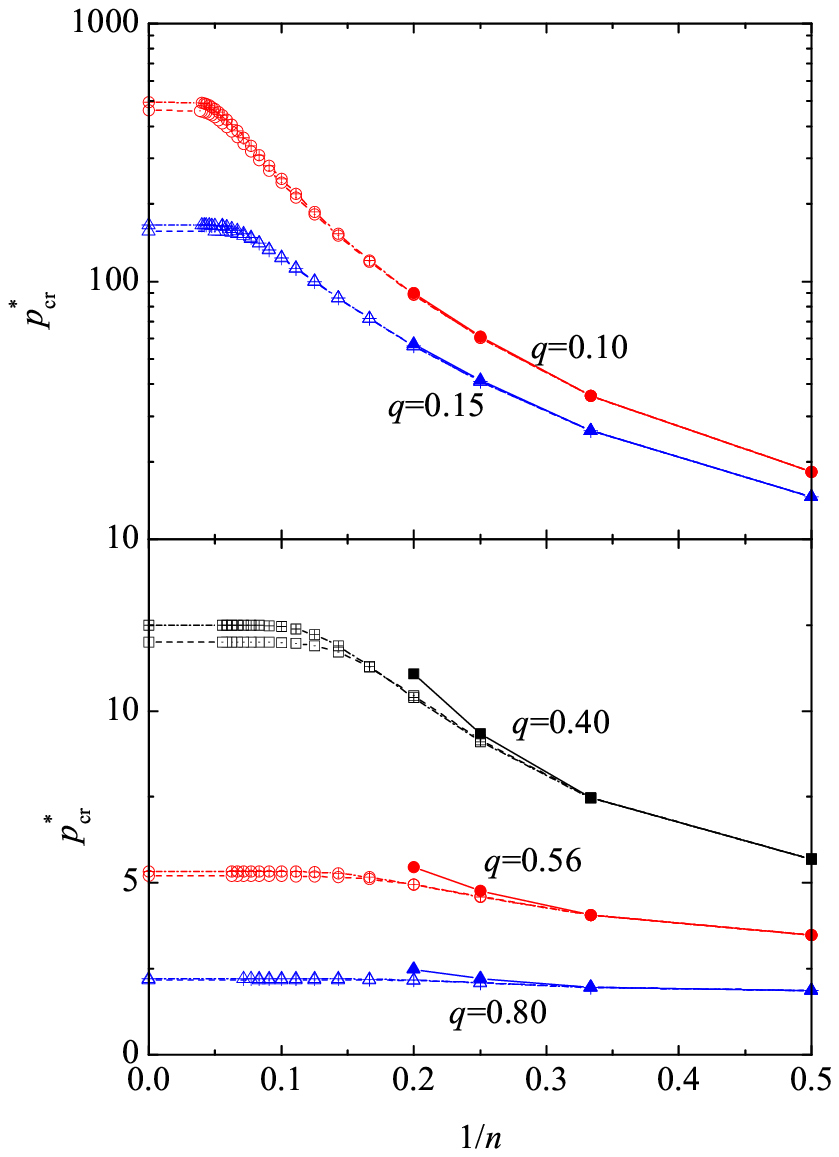}
\caption{{Same as in Fig.\ \protect\ref{fig1}, but for the critical reduced pressure ($p_{\text{cr}}^*$). No MC simulation data  are available for this quantity.}}
\label{fig3}
\end{figure}

The same information for the  critical {parameters}  $\eta_{c,\text{cr}}$,   $\eta_{p,\text{cr}}^r$,   and $p_{\text{cr}}^*$ is presented in Figs.\ \ref{fig1},  \ref{fig2}, {and \ref{fig3}}, respectively, for $q=0.10$ and $0.15$ (top panels) and $q=0.40$, $0.56$, and $0.80$ ({bottom}
panels).
In  {Figs.\ \ref{fig1} and \ref{fig2}} we have also included (at $1/n = 0$) the values obtained for {$\eta_{c,\text{cr}}$ and  $\eta_{p,\text{cr}}^r$, respectively,}  from MC simulations in the grand canonical ensemble.\cite{VH04a,VH04b,VHB05,VVPR06,FSD08,ZVBHV09,AWRE11} Also included in  {Figs.\ \ref{fig1}--\ref{fig3}} are the critical constants that one gets from the truncation at different orders of the virial expansion of two analytical equations of state for the AO model: the {FV} one by Lekkerkerker {et al.}\cite{LPPSW92} and the one proposed by three of us\cite{SHY05,SHY10} {(here denoted by the acronym SHY). The critical constants obtained from the full, nontruncated analytical equations of state are  represented in Figs.\ \ref{fig1}--\ref{fig3} at $1/n=0$. For the sake of completeness, we display below the FV and SHY equations of state:}
 \bal
{Z_{\text{FV}}=}&{x_1 Z_{\text{CS}}(\eta_c)+\frac{x_2}{1-\eta_c}+\frac{x_2 q \eta_c}{(1-\eta_c)^2}\Big[3+3q+q^2}
 \nn
   &{+3q(3+2q)\frac{\eta_c}{1-\eta_c}+9q^2\frac{\eta_c^2}{(1-\eta_c)^2}\Big],}
   \label{FV}
  \eal
 \bal
{Z_{\text{SHY}}=}&{x_1 Z_{\text{CS}}(\eta_c)+\frac{x_2}{1-\eta_c}+x_2 q\left\{\left(1-5q-\frac{11}{3}q^2\right)\right.}
 \nn
   &{\left.\times \frac{ \eta_c }{1-\eta_c}+\frac{1}{2}\left(1+4q+\frac{7}{3}q^2\right)\left[Z_{\text{CS}}(\eta_c)-1\right]\right\},}
   \label{shy}
  \eal
{where  $Z_{\text{CS}}(\eta)=(1+\eta+\eta^2-\eta^3)/(1-\eta)^3$ is the Carnahan--Starling compressibility factor of a {one-component} HS fluid.}
It should be pointed out that these two equations of state yield \emph{only} the exact second and third virial coefficients.
{Despite the similarity of Eqs.\ \eqref{FV} and \eqref{shy}, they are derived from quite different routes. The FV theory is specifically constructed for the colloid-polymer AO model, while the SHY approach extends to any HS  mixture (additive or not) with any number of components in any dimensionality. In the FV theory\cite{LPPSW92,BVS14} the free energy of the system is expressed as a sum of a term corresponding to a pure colloidal
suspension in the volume $V$  and a term corresponding
to a pure polymer solution in the volume $\alpha(\eta_c)V$, where the free volume fraction $\alpha(\eta_c)$ is motivated by scaled particle theory. On the other hand, in the SHY approximation\cite{SHY05,SHY10} the excess free energy of the mixture is written as a linear combination of that of a pure system and $-Nk_BT\ln(1-\eta)$ with coefficients such that the second and third virial
coefficients are reproduced.}

One immediately sees {from Table \ref{tab2} and Figs.\ \ref{fig1}--\ref{fig3}} that the exact virial expansion yields a very slow convergence and that already when the series is truncated after the fourth virial coefficient, the prediction for $\eta_{c,\text{cr}}$  is higher than the MC value irrespective of the size ratio {(see Fig.\ \ref{fig1})}. The same happens with the truncated virial expansions of the {FV and SHY}  equations of state. These, however, reach a maximum at a given level of truncation and then slowly decay to their final (convergent) value, which again is higher than the MC value, much more when $q \le 0.15$ and with reasonable accuracy for both $q=0.56$ and $q=0.8$. As far as {Fig.\ \ref{fig2}}  is concerned, we see that for the small size ratios $q \le 0.15$, one can hardly distinguish between the predictions {for $\eta_{p,\text{cr}}^r$} of {the FV and SHY}  equations of state and those of the exact virial series up to the fifth virial coefficient, and that the $n \to \infty$  limit prediction for both equations of state compares reasonably well with the MC value. For  higher size ratios, again the predictions of the exact virial series and those that follow from the analytical equations of state differ after the truncation at the level of the third virial coefficient, and the predictions of the full equations of state underestimate the MC values.
{Less definite conclusions can be extracted from Fig.\ \ref{fig3} due to the absence of reported MC values for the critical pressure. In any case, from the behavior of the analytical equations of state one can infer that the convergence  of $p_{\text{cr}}^*$ with increasing $n$ becomes much smoother as the size ratio increases.}

\section{Concluding remarks}
\label{sec4}

In view of the results of {Sec.\ \ref{sec3}}, a few comments are in order. To begin with, we first provided analytical expressions for the virial coefficients (up to the fourth) of the AO model and then computed numerically the fifth one for selected values of the size ratio $q$. With this input, by truncating the virial expansion after the $n$th term, where $n$ goes
from $2$ to $5$, we computed the critical constants of this model for a few values of $q$. The convergence we obtained for the final value of these constants is generally slow, especially for $q \le 0.15$.

The comparison with the MC values results is encouraging in the case of {the critical reservoir polymer packing fraction}
$\eta_{p,\text{cr}}^r$ but not in the case of the critical colloid packing fraction {$\eta_{c,\text{cr}}$}. In fact, in the former case, the extrapolation to $n \to \infty$ of the estimates obtained from the truncated virial expansions seem to be consistent with the simulation data.
In the case of $\eta_{c,\text{cr}}$, on the other hand, as one adds one more virial coefficient (what should in principle lead to better results) the observed trend is the increase of the value of the critical packing fraction and already with three ($q =0.1$, $0.05$, $0.4$, {$q = 0.56$}) or four ($q =0.80$) virial coefficients such value is above the simulation one.

The inconsistency between the estimated and simulation values of
$\eta_{c,\text{cr}}$ suggests two possible scenarios: either (a) the addition of more terms in the virial expansion will reverse the observed tendency {(and  the limit $n\to\infty$ would eventually be consistent with the MC value)} or (b) the present approach presents a serious limitation for studying fluid-fluid demixing {(for instance, if the critical density is beyond the radius of convergence of the virial series)}.
{Notice that the coexistence curve in the $\eta_c$ vs $x$ plane as obtained from MC simulations\cite{VH04a} is  very flat, so determining the precise location of the critical point is not an easy task. Under those circumstances and maybe others related to the difficulty of estimating the critical points with grand canonical simulations, as discussed in Ref.\  \onlinecite{KF04}, it is conceivable that the reported MC values of $\eta_{c,\text{cr}}$ may not be extremely accurate.}
On the other hand, the analysis of the truncated virial expansion that follows from two analytical equations of state indicates that the first scenario is indeed possible but elucidation of this issue  certainly requires further research. In any event, the usefulness of the virial approach to the demixing problem cannot be ruled out at this stage.

\begin{acknowledgments}
M.L.H., A.S., and S.B.Y. acknowledge the financial support of the Spanish Government through Grant No.\ FIS2013-42840-P and  the Junta de Extremadura (Spain) through Grant No.\ GR10158 (partially financed by FEDER funds). Thanks are also due to Prof.\ Bob Evans for suggesting this problem to us and for rich and interesting discussions and a very fruitful exchange of correspondence.

\end{acknowledgments}

\end{document}